\documentclass[aps,amssymb,prb]{revtex4}
\usepackage{epsf,epsfig}
\begin{document}
\title{Weak Localization and Integer Quantum Hall Effect in a
  Periodic Potential}

\author{G. Schwiete$^1$, D. Taras-Semchuk$^2$,  and K. B. Efetov$^{1,3}$}
\address{
$^1$Theoretische Physik III, Ruhr-Universit\"{a}t
Bochum, Universit\"{a}tsstr. 150, 44780 Bochum, Germany\\
$^2$Theory of Condensed Matter, Cavendish Laboratory, Madingley Road, Cambridge CB3 OHE, UK \\
$^3$L. D. Landau Institute for Theoretical Physics, 117940 Moscow, Russia
}

\date{\today}
\begin{abstract}
We consider magnetotransport in a disordered two-dimensional
electron gas in the presence of a periodic modulation in one
direction. Existing quasiclassical and quantum approaches to this
problem account for Weiss oscillations in the resistivity tensor
at moderate magnetic fields, as well as a strong
modulation-induced modification of the Shubnikov--de Haas
oscillations at higher magnetic fields. They do not account,
however, for the operation at even higher magnetic fields of the
integer quantum Hall effect, for which quantum interference
processes are responsible. We introduce a field theory approach,
based on a nonlinear $\sigma$ model, which encompasses naturally
both the quasiclassical and quantum mechanical approaches, as well
as providing a consistent means of extending them to include
quantum interference corrections. A perturbative
renormalization-group analysis of the field theory shows how weak
localization corrections to the conductivity tensor may be
described by a modification of the usual one-parameter scaling,
such as to accommodate the anisotropy of the bare conductivity
tensor. We also show how the two-parameter scaling, conjectured as
a model for the quantum Hall effect in unmodulated systems, may be
generalized similarly for the modulated system. Within this model
we illustrate the operation of the quantum Hall effect in
modulated systems for parameters that are realistic for current
experiments.
\end{abstract}

\maketitle

\section{Introduction}
The problem of electron motion in a disordered conductor in a
periodic potential and strong magnetic field displays a rich
combination of phenomena of both classical and quantum origin. The
complexity arises from the appearance of two independent types of
periodicities, of the potential and of the cyclotron orbits, which
may interplay in complicated ways. Perhaps the most striking
effect is known as Weiss oscillations\cite{Weiss},  whereby very
large oscillations in the resistivity are induced by even a weak
periodic potential at moderate magnetic fields. At different
magnetic fields or temperatures, other types of contributions to
the resistivity become significant. For example, at very small
fields, a positive magnetoresistance results from a classical
mechanism of channeling of orbits\cite{Beton,Streda,Menne}. At
larger magnetic fields, the resistivity develops Shubnikov--de
Haas oscillations, which originate from the onset of Landau
quantization and may still be affected by the periodic
potential\cite{Weiss,GWK,Winkler,Z+G,P+V}. At even higher fields,
the integer quantum Hall effect\cite{Klitzing} (IQHE) becomes
operative due to the contribution of quantum interference
processes.

Much of the above phenomenology has been thoroughly analyzed
theoretically, by a variety of techniques including quantum
mechanical approaches\cite{GWK,Winkler} involving
diagrammatics\cite{Z+G,Manolescu,Gross} or solution of a quantum
Boltzmann equation\cite{P+V}, and quasiclassical
approaches\cite{Beenakker,M+Wolf,Beton,Streda,Menne}. The
experimental realization of these systems is also well
advanced\cite{Weiss,GWK,Winkler,Beton,Geim,Milton} with precise
confirmation of the theoretical predictions already possible. So
far however a theory for the influence of quantum interference
processes in a periodic potential and high magnetic field has been
missing. Such processes lead to weak localization corrections to
the conductivity which become significant at low temperatures (see
e.g. Refs.\cite{Altshuler,Bergmann,Chakravarty,Belitz} for
reviews). At high magnetic fields, they are also responsible for
the operation of the IQHE\cite{Klitzing,Pruisken}, whereby the
Hall conductivity becomes quantized at low temperatures. At the
same time, the observation of the IQHE in systems with a weak
potential modulation is well within current experimental
capability\cite{Geim,Bagwell}.

This paper aims to fill this gap by showing how the standard
theory for weak localization and the IQHE may be generalized so as
to incorporate the presence of a periodic potential. We employ a
field-theory approach based on a nonlinear $\sigma$
model\cite{Wegner,ELK,Efe97,Pruisken,Levine} which is well
established in the study of mesoscopic disordered conductors. In
deriving the appropriate field theory, we are able to show how
previous theoretical calculations of the conductivity, within both
quasiclassical\cite{Beenakker,Menne} and quantum mechanical
approaches\cite{Z+G}, may be recovered in a natural way within the
field theory formalism. This enables us to extend previous
theoretical results in a consistent way so as to include the
influence of quantum interference processes.

The experiments of Weiss {\em et al}.\cite{Weiss} employed weakly
modulated two-dimensional (2D) electron systems of high mobility
with a well-known period, $a \sim 300\;{\rm nm}$, much less than
the mean free path, $\ell \sim 10\;\mu {\rm m}$.  Such samples
were engineered using a holographic modulation technique, based on
the persistent photoconductivity effect in GaAs/Al$_x$Ga$_{1-x}$As
heterostructures at low temperatures $\sim 4.2\;{\rm K}$. The
Weiss oscillations appear in only one component of the resistivity
tensor, $\rho_{xx}$, when the modulation is in the $x$ direction.
Furthermore they appear at magnetic fields $B$ such that the
cyclotron radius $R_c = v_F/\omega_c$ (where $\omega_c=eB/mc$ is
the cyclotron frequency, $-e$ is the electron charge and $v_F$ is
the Fermi velocity) satisfies the commensurability condition
 $2R_c = (n-1/4)a$, for integer
$n$. Hence the oscillations are periodic in $1/B$. In addition they
are relatively stable with respect to temperature, suggesting a
quasiclassical origin.

At even higher magnetic fields, such that the cyclotron radius is much
less than the period of modulation ($R_c \ll a$),
the quasiclassical theory  predicts that the $\rho_{xx}$ component
shows a
large, nonoscillatory increase
proportional to $B^2$, leading to a strong positive
magnetoresistance. This result has been confirmed
in experiments\cite{Geim} for which the temperature was kept
deliberately high
so as to avoid the intervention of the IQHE.

A limitation of the quasiclassical approach is that it fails to
account for the renormalization of single-particle properties,
such as the density of states and scattering lifetime, by the
strong magnetic field according to Landau quantization. Even in
unmodulated samples, quantization leads to oscillations in the
density of states with respect to magnetic fields, and hence to
Shubnikov--de Haas oscillations in the resistivity. In samples
with the periodic potential, the Shubnikov--de Haas oscillations
start to appear at higher magnetic fields than the Weiss
oscillations (for $R_c$ an integer multiple of the Fermi
wavelength $\ll a$). A quantum-mechanical approach that does allow
for such quantization effects has been provided by Zhang and
Gerhardts\cite{Z+G} (see also Peeters and Vasilopoulos\cite{P+V});
it is a diagrammatic treatment that generalizes the approach of
Ando and co--workers\cite{Ando} to modulated samples. The
quantum-mechanical approach is then capable of describing both
Weiss oscillations and the Shubnikov--de Haas oscillations (also
affected by the modulation) at higher magnetic fields.

In principle, the calculation of weak localization corrections in
a strong magnetic field (unitary ensemble) is possible even in the
presence of a periodic potential, by a generalization of the
diagrammatic approach of Zhang and Gerhardts\cite{Z+G}, but the
procedure would be complicated (although the simpler orthogonal
case has been examined diagrammatically for a periodic magnetic
field\cite{Shelankov,Wang}).  Instead, the calculation of
high-order diagrams is more convenient in the field-theory
formalism and, furthermore, with this method the possibility
exists of calculating contributions of diagrams to all orders by
the renormalization-group technique. We show how the field-theory
takes the form of a nonlinear $\sigma$ model with a topological
term~\cite{Wegner,ELK,Efe97,Pruisken,Levine}. The effective
Lagrangian is slightly nonstandard since it contains an anisotropy
in the coefficients, corresponding to the difference between the
longitudinal conductivities in the $x$ and $y$ directions,  due to
the periodic potential.

The effect of weak localization corrections to the conductivity
for unmodulated samples is accounted for by a scaling (one-parameter scaling) of the conductivity with the system size.
As a first step we derive the analog of the one-parameter scaling
for the conductivity\cite{Wegner,ELK,Hikami,Efe97} in the
modulated system by means of a perturbative RG analysis of the
effective Lagrangian. We then turn to the study of the IQHE,
implementing a generalization of a two-parameter scaling, which
has been conjectured\cite{Levine,Khmelnitskii,Pruis2} as a model
for the IQHE in unmodulated systems. We examine how the
resistivity tensor should be affected at low temperatures by the
IQHE in modulated samples for parameters that are realizable in
actual experiments. We see, for example, how the Hall conductivity
becomes quantized under scaling at low temperatures, while in the
regions between the plateaus the longitudinal conductivities
develop peaks of differing heights
 according to the anisotropy in
the $x$ and $y$ directions.

{\em Model.} In the following, we employ the Hamiltonian for the
disordered conductor in a magnetic field and periodic potential in
two dimensions:

\begin{eqnarray}
H&=&H_0 - V({\bf r}), \nonumber\\
H_0&=&
\frac{1}{2m}\left({\bf p}-\frac{e {\bf A}}{c}\right)^2
+U_0 \cos(qx),
\label{Hamiltonian}
\end{eqnarray}

Here ${\bf A}$ is the vector potential, so that $\nabla\times {\bf
A} = (0,0,B)$, where $B$ is the perpendicular, uniform magnetic
field. Also $U_0 \cos(qx)/(-e)$ is the periodic potential which is
taken to be weak ($U_0\ll \epsilon_F$, where $\epsilon_F$ is the
Fermi energy) and with modulation period $a= 2\pi/q$. $V$ is the
disorder potential, which we assume to be $\delta$-correlated in
space, with the associated scattering time $\tau$. We assume that
the mean free path $\ell=v_F\tau$ greatly exceeds the modulation
period $a$.

The plan for the remainder of the paper is as follows. In Sec.~
\ref{sec:sigma} we describe the field theory approach and derive
the effective Lagrangian for the system. In Sec.~
\ref{sec:scaling} we show how the field theory provides the
scaling of the conductivity tensor under changes of length scale
due to the contribution of quantum interference processes, and
hence a description of the IQHE in these samples. Section
\ref{sec:summary} concludes with a summary and discussion.

%******************************************

%* Field Theory Approach                  *

%******************************************

\section{Field-Theory Approach}
\label{sec:sigma}

In this section we describe a field-theoretical approach to the
description of a disordered conductor in the presence of a
periodic potential and a strong magnetic field. The approach is
based on
 a diffusive nonlinear $\sigma$ model, which will
serve as a tool for the analysis of quantum interference effects in the remainder of this paper. In section \ref{mainfindings} we discuss this model and the relation to previous semiclassical\cite{Beenakker} and quantum mechanical\cite{Z+G} approaches. Technical details of the derivation are presented in section \ref{derivation}.

\subsection{Description of field theory approach}
\label{mainfindings}
 The field-theory apparatus that
we employ is by now well established in the study of spectral and
wave-function statistics of disordered conductors. It is based on
a functional integral expression of electron Green functions in
the presence of disorder, and in the diffusive regime takes the
form of a nonlinear $\sigma$ model in terms of a $Q$
matrix\cite{Wegner,ELK,Efe97}. In two dimensions, the field-theory
technique provides a convenient means of calculating weak
localization corrections to the conductivity, a task that may
become very cumbersome by the more conventional technique of
diagrammatics\cite{GLK,Hikami}. The field-theory may also provide
a resummation of diagrams to all orders by a renormalization-group
procedure\cite{Wegner,ELK,Hikami,Efe97}.

We show below that the effective Lagrangian for the disordered
conductor in the presence of a periodic potential and strong
magnetic field takes the form
\begin{eqnarray}
\label{diffaction}
{\cal L}[Q]=\frac{1}{16} \int dr\, {\rm Str}
\left\{\sigma^0_{xx}(\nabla_x
Q)^2+\sigma^0_{yy}(\nabla_y
Q)^2-\sigma^0_{xy}\tau_3Q [\nabla_x Q,\nabla_y Q]\right\},
\end{eqnarray}
where $\sigma^0_{ij}$ is the classical conductivity tensor of the system, in units of $e^2/h$.
The 8$\times$8 supermatrix field $Q({\bf r})$ satisfies the nonlinear constraint $Q({\bf r})^2=1$. The supertrace operation is defined in Ref.~\cite{Efe97}.

In the absence of a strong magnetic field or modulation, a
renormalization-group analysis of the first two terms in
Eq.~(\ref{diffaction}) leads to the well-known one-parameter
scaling\cite{Wegner,ELK,Hikami,Efe97} of the conductivity with
system size due to weak localization.

The final term in Eq.~(\ref{diffaction}) is known as a topological term and appears in the theory of the IQHE proposed by Pruisken and
collaborators\cite{Pruisken,Levine}. The influence of the extra term
does not appear within perturbation theory, but becomes evident only
through a nonperturbative analysis. Such an analysis
 has been
conjectured\cite{Levine,Khmelnitskii,Pruis2} in the form of a
two-parameter renormalization-group procedure.  The two parameters
are the longitudinal and Hall conductivities that follow coupled
scaling equations with respect to changes of length scale.

While the validity of the two-parameter scaling has been
vigorously debated (see, e.g., Refs.~\cite{Zirnbauer,Weid+Z}), it
has remained a valuable guide to experimental\cite{Wei} and
numerical\cite{Huckestein} data for a number of years. In the
following, we take a pragmatic approach by not contesting the
validity of the two-parameter scaling and its relation to the
proposed field theory. Instead, we assume its validity and explore
how it may be generalized to take account of the periodic
potential.

The main difference of the Lagrangian (\ref{diffaction}) from the unmodulated
 case is the anisotropy in the
diffusion coefficients for the $x$ and $y$ directions. The
Lagrangian, therefore, contains three, rather than two parameters,
which scale together under changes of length scale. A simple scale
transformation, however, maps the Lagrangian to an isotropic
version for which the usual two-parameter scaling may be applied.

This Lagrangian applies in the diffusive limit, that is, to
configurations of the $Q({\bf r})$ field that  vary on scales much
longer than the mean free path. This allows the calculation of the
contribution of low-momentum relaxational modes to weak
localization corrections to the conductivity. Although momentum
relaxation is diffusive on large length scales, electron motion on
the scale of the periodic potential is ballistic, since the
modulation period is much less than the mean free path. To arrive
at the above Lagrangian it is therefore necessary to integrate
over degrees of freedom corresponding to length much smaller than
the mean free path.

One way to do so is to start from a description of ballistic
electron motion on length scales much less than the mean free
path. Such a description is provided by the "ballistic $\sigma$
model"\cite{Muz95,Andreev} which, as the name suggests,
generalizes the diffusive $\sigma$ model to the ballistic regime.
Starting from this model we show in Sec.~\ref{derivation} how the
contribution from short length scales may be integrated out. The
result is the Lagrangian of a diffusive $\sigma$ model that
describes the interaction of diffusion modes on scales longer than
the mean free path.

When derived in this way, the effective Lagrangian takes the form
of Eq.~(\ref{diffaction}), but with the $\sigma_{ij}^0$
coefficients replaced by $\sigma_{ij}^{\rm qc}(h/e^2)$. Here
$\sigma^{qc}_{ij}$ correspond precisely to the components of the
quasiclassical conductivity tensor as derived from the Boltzmann
equation for the modulated system \cite{Beenakker}. One finds
that, in the quasiclassical approach, only the xx component of the
quasiclassical resistivity tensor is affected by the modulation,
all other components being the same as in the unmodulated case, so
that $\rho^{\rm qc}_{xx}=\rho_0= (2e^2\nu D)^{-1}$ and $\rho^{\rm
qc}_{xy}=-\rho^{\rm qc}_{yx}=\omega_c \tau \rho_0$ (here
$D=v_F^2\tau/d$ is the diffusion coefficient in dimension $d$).
For details of the solution of the Boltzmann equation, we refer
the reader to Beenakker\cite{Beenakker}. Here we only remark that
for moderate magnetic fields, weak enough that the cyclotron
radius is much larger than the modulation period ($R_c  \gg a$),
but strong enough that $\omega_c\tau \gg 1$, the quasiclassical
result simplifies to

\begin{eqnarray}
\frac{\rho^{\rm qc}_{xx}}{\rho_0} = 1+\frac{1}{2\pi}
\left(\frac{U_0}{\epsilon_F}\right)^2
(\omega_c\tau)^2 R_c q \cos^2\left(R_c q-\frac{\pi}{4}\right),
\qquad \qquad(R_c q \gg 1).
\label{Been2}
\end{eqnarray}
Equation.~(\ref{Been2}) demonstrates the oscillatory dependence of
the resistivity known as Weiss oscillations. For strong magnetic
fields, such that $R_c\ll a$, a very large oscillatory increase in
$\rho_{xx}$ proportional to $B^2$ is predicted and has been
observed experimentally in, e.g., Geim et al.~\cite{Geim}.

However, for high magnetic fields ($\omega_c\tau\gg 1$), the
quasiclassical results are reliable only at sufficiently high
temperatures ($k_B T \gg \hbar/\tau$), while at lower temperatures
the effects of quantization on single-particle properties such as
the density of states must be taken into account by a
quantum-mechanical approach. These effects are not taken into
account in the derivation via the ballistic $\sigma$ model, since
here the magnetic field is treated only as a weak perturbation.
Indeed, once temperatures are low enough for weak localization
corrections to be significant ($k_B T \sim \hbar/\tau$), then
quantization is already well established (as $k_B T \ll \hbar
\omega_c$, since $\omega_c\tau \gg 1$). Hence this method of
deriving the Lagrangian is unfortunately of no use in  describing
the effects of quantum interference processes at high magnetic
fields.

An improvement on this situation may be found by proceeding
instead along a second route to derive the effective Lagrangian,
following more closely the original lines of
Pruisken\cite{Pruisken}. He showed how to derive a $\sigma$ model
in the presence of a strong magnetic field, including the effects
of quantization. The Lagrangian is applicable at high values of
the Landau-level index $n$, such that the fluctuations of the
density of states may be neglected. We show below in
Sec.~\ref{derivation} how this method may be adapted to describe a
disordered conductor in a strong magnetic field and a periodic
potential. This route bypasses the intermediate step of a
ballistic $\sigma$ model, instead computing more directly the
final diffusive Lagrangian.

\begin{figure}[tbp]
\begin{center}
\epsfig{file=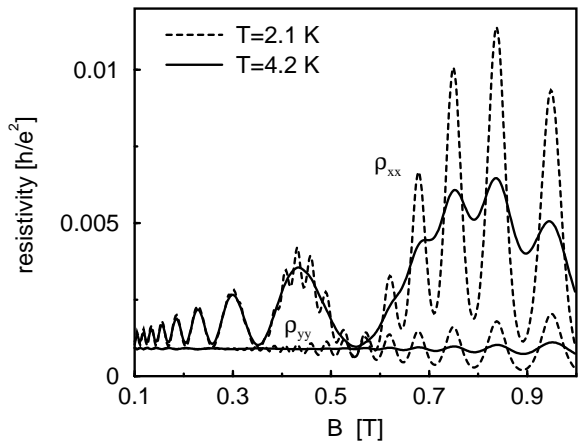,height=6cm}
\caption{\label{fig:RESPV} Resistivity components as a function of magnetic field $B$ for two different temperatures using the approximation scheme of Peeters and Vasilopoulos \cite{P+V}. Corrections to the Hall resistivity are small in this regime \cite{Z+G,P+V}. Parameters are $a=$200 nm, $U_0=$0.3 meV, $n_{el}=$3.4$\times$10$^{11}$ cm$^{-2}$, $\hbar/\tau=$0.011 meV.}
\end{center}
\end{figure}

While the resulting Lagrangian takes again the form displayed in
Eq.~(\ref{diffaction}), now the $\sigma_{ij}^0$ coefficients
correspond precisely to $\sigma_{ij}^{\rm qu} h/e^2$, where
$\sigma_{ij}^{\rm qu}$ are the components of the conductivity
tensor calculated in the fully quantum-mechanical approach of
Ref.~\cite{Z+G}.
 Since the effects of
quantization are included this approach correctly describes
certain features that are observed in experiment but are beyond
the quasiclassical approach. Some typical results for the
resisitivity components ($\rho^{\rm qu}_{ij}$) calculated within
the quantum-mechanical approach are shown in Fig.~\ref{fig:RESPV}.

%***************************************************

%* Figure: RESPV                                    *

%***************************************************

As well as the appearance of Weiss oscillations in $\rho^{\rm
qu}_{xx}$ at moderate magnetic fields (up to $\sim 0.4\;T$ in
Fig.~\ref{fig:RESPV}), we see that both $\rho^{\rm qu}_{xx}$ and
$\rho^{\rm qu}_{yy}$
 show strong, in-phase
Shubnikov--de Haas oscillations at higher magnetic fields. These
latter oscillations reflect those of the density of states. At
weaker fields, for which the Weiss oscillations in $\rho^{\rm
qu}_{xx}$ are visible, weaker, out-of-phase oscillations in
$\rho^{\rm qu}_{yy}$ also appear. Again, the latter oscillations
in $\rho^{\rm qu}_{yy}$ are beyond the quasiclassical approach.

 The quasiclassical
result (\ref{Been2}) for the change in $\rho^{\rm qc}_{xx}$ (for
$\omega_c\tau \gg 1$ and $R_c q \gg 1$) may be reproduced by the
quantum mechanical approach, as long as the temperature is high
enough for the thermal broadening to be greater than the
separation of the Landau levels, $k_B T \gg \hbar \omega_c$ (see,
e.g., Refs.\cite{Z+G,P+V}). As the temperature is lowered,
however, the region of applicability of
 the quasiclassical formula (\ref{Been2}) shrinks to progressively
lower magnetic
 fields: The sinusoidal Weiss
 oscillations start to interfere at
low temperatures with the Shubnikov--de Haas oscillations. Indeed,
at vanishingly low temperatures, the quasiclassical result becomes
unjustified for all magnetic fields.

The  Lagrangian (\ref{diffaction}), supplied with the values of
the conductivities determined by the quantum-mechanical approach,
may then be used for a reliable description of quantum
interference processes at high magnetic fields, and hence the
operation of the IQHE.

\subsection{Derivation of the Field Theory}

\label{derivation}

In this section we describe in more detail the derivation of the
effective Lagrangian (\ref{diffaction}). In the field-theory
approach, it is necessary to average the functional integral
expression for products of Green functions over the disorder
configurations. To perform the  disorder average, several possible
techniques exist, including  the use of either supersymmetry or
the replica trick. Although we may use either of these two
approaches, we choose here the supersymmetry approach, since it is
free of certain technical problems regarding nonperturbative
calculations that appear for the replica approach. The starting
point for both routes is a functional integral representation of
the partition function, $Z$, in terms of "superfields"
$\Psi,\bar{\Psi}$:
\begin{eqnarray}
Z &=& \int D[V] P[V] D[\Psi,\bar{\Psi}] \exp {-\cal L}(\Psi,\bar{\Psi},V), \\
{\cal L} &=& i \int dr  \bar{\Psi}({\bf r}) (E-H-i\delta\Lambda)\Psi({\bf r}).
\label{Lag1}
\end{eqnarray}

Here the superfields $\Psi,\bar{\Psi}$ contain eight components,
corresponding to fermion/boson, retarded/advanced and
time-reversal sectors~\cite{Efe97}. Also $\Lambda ={\rm
diag}(1,-1)$ in advanced-retarded space space, $\delta$ is a
positive infinitesimal and $H$ is the Hamiltonian
(\ref{Hamiltonian}) up to the substitution ${\bf A}\rightarrow
\tau_3 {\bf A}$, where $\tau_3=$diag$(1,-1)$ in time-reversal
space. The short-range random potential $V({\bf r})$ is taken to
be Gaussian distributed, according to

\begin{eqnarray*}
P[V({\bf r})] = \frac{1}{{\cal Z}} \exp\left\{
-\frac{\pi\nu\tau}{\hbar}\int dr V({\bf r})^2\right\},
\end{eqnarray*}
where ${\cal Z}$ provides the normalization.

{\it Ballistic $\sigma$ Model.} The first route is to derive a
ballistic $\sigma$ model from the Lagrangian in Eq.~(\ref{Lag1}).
Andreev {\em et al}.~\cite{Andreev} have suggested a derivation of
the ballistic $\sigma$ model by means of an energy averaging
procedure, leading to a Lagrangian of the same form as that
originally proposed by Muzykantskii and Khmelnitskii\cite{Muz95}.
Following the lines of Andreev {\em et al}.\cite{Andreev}, one
treats the disorder field $V$ as a perturbation in
Eq.~(\ref{Lag1}). Performing an energy averaging on the clean
Hamiltonian leads to a term in the Lagrangian that is nonlocal and
quartic in the $\Psi$, $\bar{\Psi}$ fields. The quartic term is
then decoupled by a Hubbard-Stratonovich field, $Q({\bf r}_1,{\bf
  r}_2)$, and
the $\Psi$, $\bar{\Psi}$ fields are integrated out.

The resulting Lagrangian is then simplified by subjecting it to a
saddle point analysis, and performing a gradient expansion around
the saddle point. Note that, since at this stage we are neglecting
the effects of quantization on single-particle properties, we do
not include the renormalization of this saddle point solution by
the magnetic field, instead we treat the magnetic field as a
perturbation.

Following a Wigner transform, and after applying a
semiclassical\cite{note:semi} approximation, the $Q$ matrix
becomes a function of position, ${\bf r}$, and the direction of
momentum, ${\bf n}$, where ${\bf n}$ is a unit vector. We do not
repeat these steps here but refer to Refs.~\cite{Andreev,Muz95}
for further details. The resulting Lagrangian (ballistic $\sigma$
model) corresponding to the Hamiltonian (\ref{Hamiltonian}) is

\begin{eqnarray}
{\cal L}[Q_{\bf n}]=\frac{\pi\nu \hbar}{4} \int  dr
  {\rm Str}\left\{-\frac{1}{2\tau}\int
\frac{dn}{2\pi}
\frac{dn'}{2\pi}Q_{\bf n}({\bf r})Q_{{\bf n}'}({\bf r})-2\int
   \frac{dn}{2\pi}\Lambda\tau_3 T_{\bf n}^{-1}{\mathcal L}^0({\bf n},{\bf r})T_{\bf n}\right\},
\label{freeb}
\end{eqnarray}
In this equation ${\mathcal L}^0$ is the Louiville operator defined by
\begin{eqnarray}
{\mathcal L}^0=v({\bf r}){\bf n}.\nabla+\omega_c\partial_\phi
-\sin{\phi}v'(x)\partial_\phi
\label{Liouville}
\end{eqnarray}
where $\phi$ is the polar angle of ${\bf n}$, $\nu$ is the density
of states per spin direction at the Fermi level, and
\begin{eqnarray*}
v({\bf r})=v_F\left(1-\frac{U_0}{E_F}\cos(qx)\right)^{1/2}
\end{eqnarray*}
is the electron velocity at the Fermi surface. The supermatrix
$Q_{\bf n}({\bf r})$ satisfies the nonlinear constraint $Q_{\bf
n}({\bf r})^2=1$, as a result of which it may be parametrized by
\begin{eqnarray*}
Q_{\bf n}({\bf r})=T_{\bf n}({\bf r})\Lambda T_{\bf n}^{-1}({\bf r}).
\end{eqnarray*}
Due to the presence of the magnetic field, the $Q_{\bf n}$ field
satisfies unitary symmetry, that is, $[Q_{\bf n},\tau_3]=0$.
The $Q_{\bf n}({\bf r})$ field also satisfies the further
standard symmetries $\bar{Q}_{\bf n}=Q_{\bf n}=K Q_{\bf n}^\dagger K$, where
$\bar{Q}_{\bf n}\equiv CQ_{\bf n}^T C^T$. The matrices $C$
and $K$, as well as supertrace operation,
are defined in Ref.~\cite{Efe97}.

The Lagrangian (\ref{freeb}) is in its current form too complicated
for our purposes: it describes fluctuations of $Q_{\bf n}({\bf r})$ on
length scales smaller than the mean free path
(and longer than the Fermi wavelength) and for any
dependence on ${\bf n}$. At the same time, electron motion on length scales
much longer than the mean free path is diffusive. The
strategy therefore is to integrate out the modes corresponding to
fluctuations on length scales smaller than the mean free path, to
produce a Lagrangian  that describes the interaction of
diffusion modes on larger length scales.

To perform the integration, we isolate the the fluctuations of
lowest mass (termed massless), corresponding to matrices $Q_{\bf
n}({\bf r})$ that are independent of ${\bf n}$ and vary more
slowly than the mean free path, $\ell$. We then integrate out all
other fluctuations that preserve $Q_{\bf n}^2=1$.
 A similar procedure
has been carried out by W\"olfle and Bhatt\cite{Woe84} to derive the
diffusive Lagrangian for a disordered conductor with anisotropic
masses in the $x$ and $y$ directions (see also Refs.\cite{Aronov,TE}
for further examples).

To perform integration over the weakly massive modes, we write

\begin{eqnarray*}
Q_{\bf n}=U({\bf r}) Q^0_{\bf n}({\bf r})\bar{U}({\bf r}).
\end{eqnarray*}

Here the $U({\bf r})$ matrix, obeying the symmetries $ \bar{U}U=1$
and $\bar{U}=KU^\dagger K$, represents the massless fluctuations,
while $Q_{\bf n}^0({\bf r})$ contains the weakly massive
fluctuations. The matrix $Q_{\bf n}({\bf r})$ may, in turn, be
parametrized by its generator $P_{\bf n}({\bf r})$, for example,
by\cite{Efe97}

\begin{eqnarray*}
Q_{\bf n}^0=\Lambda\left(\frac{1+iP_{\bf n}}{1-iP_{\bf n}}\right).
\end{eqnarray*}

Here $P_{\bf n}$ is off-diagonal in retarded-advanced space
and satisfies the symmetry $P_{\bf n}= -\bar{P}_{{\bf n}}$.
Since the fluctuations represented by the $Q_{\bf n}$ matrix
are weakly massive, it
is sufficient to treat them within a Gaussian approximation:
$Q_{\bf n}=\Lambda(1+2iP_n-2P_n^2\dots)$.
Inserting into the Lagrangian (\ref{freeb}), we find

\begin{eqnarray*}
{\cal L}&=&\frac{\pi\nu\hbar}{2}
\int dr {\rm Str}\left\{-\int \frac{dn}{2\pi} \Lambda\tau_3
P_{\bf n} {\mathcal L}^0({\bf r},{\bf n})
P_{\bf n}
+\frac{1}{\tau}\int \frac{dn}{2\pi} \frac{dn'}{2\pi}
(P_{\bf n}^2-P_{\bf n} P_{{\bf n}'})
\right. \\
&&\left.
-2i\int\frac{dn}{2\pi}\Lambda \tau_3 P_{\bf n} v({\bf r}){\bf n}\cdot
{\mathbf \Phi^\perp}\right\}
\end{eqnarray*}
where ${\mathbf \Phi}\equiv\bar{U}\nabla U$, ${\mathbf
\Phi}^\perp=1/2[{\mathbf \Phi},\Lambda]\Lambda$. The next step is
to integrate over matrices $P_{\bf n}$. This amounts to a set of
Gaussian integrals with a linear term in $P_{\bf n}$, and hence
may be performed by means of a shift of $P_{\bf n}$.

The resulting Lagrangian has the form
\begin{eqnarray}
\label{diffaction2}
{\cal L}[Q]=\frac{h}{16e^2} \int dr\, {\rm Str}
\left\{\sigma^{\rm qc}_{xx}(\nabla_x
Q)^2+\sigma^{\rm qc}_{yy}(\nabla_y
Q)^2-\sigma^{\rm qc}_{xy}\tau_3Q [\nabla_x Q,\nabla_y Q]\right\},
\end{eqnarray}
It is written in terms of a matrix $Q({\bf r})\equiv U({\bf
r})\Lambda \bar{U}({\bf r})$ that depends only on position and
varies on length scales much longer than the mean free path (and
hence the modulation period, $a$). The $Q({\bf r})$ field also
satisfies the nonlinear constraint $Q({\bf r})^2=1$, as well as
the symmetries $Q=\bar{Q} = K Q^{\dagger} K$. We see that the
Lagrangian takes the usual form of a diffusive $\sigma$ model with
a topological term\cite{Pruisken}, the only non-standard feature
being the anisotropy of $\sigma_{xx}^{\rm qc}\neq \sigma_{yy}^{\rm
qc}$.

The coefficients $\sigma^{\rm qc}_{ij}$ are defined by
\begin{eqnarray}
\label{conductivities}
\sigma^{\rm qc}_{ij}=2e^2\nu
\left\langle  \int dr' \frac{dn}{2\pi} {\frac{dn'}{2\pi}}
 v({\bf r})v({\bf r}') n_i {n'}_j
\Gamma({\bf n},{\bf n}',{\bf r},{\bf r}')
 \right\rangle,
\end{eqnarray}
where the averaging in Eq.~(\ref{conductivities})
is over one period of the modulation $a$ (this averaging arises
naturally in the Lagrangian (\ref{diffaction})
since the ${\mathbf \Phi}$ matrix varies slowly on this
scale). Also, $\Gamma({\bf n},{\bf n}',{\bf r},{\bf r}')$ is defined by the
Boltzmann-like equation
\begin{eqnarray}
\left({\mathcal L}^0({\bf r},{\bf n})-\widehat{C}
\right)\Gamma({\bf r},{\bf r}',{\bf n},{\bf n}')\equiv\delta({\bf r}-{\bf
  r}') \delta({\bf n}-{\bf n}').
\label{Boltz2}
\end{eqnarray}
Here, the collision operator is given as
\begin{eqnarray*}
\widehat{C}\{F(\phi)\}\equiv-\frac{1}{\tau}\left( F(\phi)-\int
  \frac{d\phi'}{2\pi} F(\phi')\right).
\end{eqnarray*}
As we will show now the $\sigma^{\rm qc}_{ij}$ coefficients in the
Lagrangian (\ref{diffaction2}) are the components of the
quasiclassical conductivity tensor as derived by
Beenakker~\cite{Beenakker}. In his approach, the Boltzmann
equation is expressed~\cite{Beenakker} (see also
Refs.\cite{M+Wolf,Menne,Streda,Beton}) in terms of the
distribution function F({\bf r},{\bf n})
\begin{eqnarray}
\left({\mathcal L}^0-\widehat{C}\right)F({\bf r},{\bf n})=-ev({\bf r})
{\bf E}\cdot{\bf n},
\label{Boltzmann}
\end{eqnarray}
and the current ${\bf J}({\bf r})$ is given by
\begin{eqnarray}
J^i({\bf r})=-2e\nu\langle F({\bf r},{\bf n})v({\bf r}) n^i\rangle,
\label{Jdef}
\end{eqnarray}
where the averaging is over both the velocity direction ${\bf n}$
and one period of the modulation, $a$. The relation $J^i =
\sigma^{\rm qc}_{ij} E_j$ defines the quasiclassical conductivity
tensor $\sigma^{\rm qc}_{ij}$, whose inversion gives the
quasiclassical resistivity tensor $\rho^{\rm qc}_{ij}$.

Comparing Eqs.~(\ref{Boltz2}) and (\ref{Boltzmann}), we see
that $\Gamma({\bf n},{\bf n}',
{\bf r},{\bf r}')$ is related to
the distribution function
$F({\bf r},{\bf n})$ by

\begin{eqnarray*}
F({\bf r},{\bf n})
=-e\int dr' \frac{dn'}{2\pi}v({\bf r}'){\bf E}\cdot {\bf n}'
\Gamma({\bf r},{\bf r}',{\bf n},{\bf n}').
\end{eqnarray*}
Comparing also Eqs.~(\ref{conductivities}) and (\ref{Jdef}), we
see that the $\sigma^{\rm qc}_{ij}$ coefficients in the Lagrangian
(\ref{diffaction2}) calculated by this method coincide precisely
with the components of the quasiclassical conductivity as
derived\cite{Beenakker} from the Boltzmann equation
(\ref{Boltzmann}). In this way we see that we have rederived the
quasiclassical results within the field-theory formalism.

As discussed above, the inadequacy of the Lagrangian
(\ref{diffaction2}) is that it neglects the renormalization of the
single-particle properties by the strong magnetic field. At high
magnetic fields ($\omega_c\tau \gg 1$), once the temperature is
low enough for the weak localization corrections to be significant
($k_B T \sim \hbar/\tau$), the quasiclassical results for the
conductivity\cite{Beenakker} have already become unreliable due to
their neglect of quantization. In order to include these effects,
and hence account for weak localization effects reliably, we need
to follow a different route to derive the Lagrangian. Such a route
for unmodulated systems has been provided by
Pruisken\cite{Pruisken}, whose method may be adapted (as we show
here) to include a periodic potential. This method includes the
renormalization of the saddle point equation for the $Q$ matrix by
the strong magnetic field. It does not require a derivation of a
ballistic $\sigma$-model as an intermediate step, but provides
more directly
 the final form of the Lagrangian. While the first route contains certain
 parallels with the quasiclassical approach \cite{Beenakker}, the second route contains closer
 parallels with the quantum-mechanical approach\cite{Z+G}.

{\it Generalization of Pruisken derivation.} We now present the
second route to the derivation of the effective Lagrangian
(\ref{diffaction}), starting
 again from form (\ref{Lag1}) for the Lagrangian. The
approach now is to average over the short-range disorder. This
produces a term in the Lagrangian that is quartic in the $\Psi$,
$\bar{\Psi}$ fields. The quartic term may then be decoupled by a
(now local) Hubbard-Stratonovich field $Q({\bf r})$. After
integrating out the $\Psi$, $\bar{\Psi}$ fields, one finds

\begin{eqnarray}
{\cal L}[Q] = \int dr\left\{\frac{\hbar\pi\nu}{8\tau}{\rm Str}\,{Q}^2
-\frac{1}{2}{\rm Str}\,{\rm ln}
\left[-i\left(\pi_{\mu}\pi^{\mu}-E+U_0\cos(qx)\right)
+\delta\Lambda
+\frac{\hbar Q}{2\tau}\right]\right\}
\label{Trln}
\end{eqnarray}
where $\pi_{\mu}\equiv-i\hbar \nabla_\mu-e\tau_3A_{\mu}/c$.
Here Q satisfies the symmetry $Q=\bar{Q}$.

The Lagrangian (\ref{Trln}), in principle, provides an exact
description of the system, although in its current form it is too
general to be useful. Instead, one proceeds by finding the saddle
point value of $Q$ that minimizes the Lagrangian, and by
performing a gradient expansion about this minimum. The saddle
point equation may be written as

\begin{eqnarray}
Q=\frac{i}{\pi\nu}<{\bf r}|
\left(\pi_{\mu}\pi^{\mu}+U_0\cos(qx)-E+i\delta \Lambda
+\frac{i \hbar Q}{2\tau}\right)^{-1}|{\bf r}>.
\label{sad1}
\end{eqnarray}

In order to find the matrix inverse of the operator in
Eq.~(\ref{sad1}), we make use of the eigenvalues and
eigenfunctions of the operator
$H_0=\pi_{\mu}\pi^{\mu}+U_0\cos(qx)$. Using the Landau gauge ${\bf
A}=(0,Bx,0)$, one may write the eigenfunctions of $H_0$ in the
form $\psi_{k,n} (x,y) = L_y^{-1/2} \exp(i\tau_3k
y)\phi_{n,x_0}(x)$, where $L_y$ is a normalization length. The
center coordinate $x_0=l_B^2 k$ remains a good quantum number
despite modulation, where $l_B= (\hbar c/eB)^{1/2}$ is the
magnetic length. The $\phi_{n,x_0}(x)$ are eigenfunctions of the
Hamiltonian

\begin{eqnarray}
H_{x0} = -\frac{\hbar^2}{2m}\frac{d^2}{dx^2}
+\frac{1}{2}m\omega_c^2(x-x_0)^2 +U_0 \cos(q x).
\label{H0a}
\end{eqnarray}

In the absence of modulation ($U_0=0$), Eq.~(\ref{H0a}) represents
the Hamiltonian of a harmonic oscillator. We proceed by a
first-order perturbative expansion modulation $U_0$. This turns
out to be a very good approximation for typical parameters that we
consider, such as a weak periodic potential. It is very difficult
to improve on this approximation analytically, although exact
numerical diagonalizations have been performed for the density of
states\cite{GWK,Z+G}. Within the perturbative expansion, the
eigenvalues $\epsilon_n(x_0)=\epsilon_n(x_0+a)$ of the Hamiltonian
(\ref{H0a}) are given by
\begin{eqnarray}
\epsilon_n(x_0) \simeq E_n(x_0) \equiv E_n+u_n \cos(q x_0),
\label{eneq}
\end{eqnarray}
where $E_n= \hbar \omega_c (n+1/2)$ are the unperturbed Landau
energies, and $u_n = U_0 \exp\left(-1/2X\right) L_n(X)$,
where $X= q^2 l_B^2/2$ and $L_n$ are the Laguerre polynomials\cite{Abr72}.

We see that the modulation lifts the degeneracy of the Landau
levels and discrete levels are broadened into bands, whose width
depends on the band index $n$ in an oscillatory manner (due to the
behavior of the Laguerre polynomials at large $n$). It is this
oscillatory dependence that leads to the Weiss oscillations in the
resistivity. The validity of the perturbation theory depends only
on the smallness of the $u_n$ parameter, which is assured for
large values of the Landau-level index $n$. We refer the reader
to, e.g., Refs.\cite{GWK,Winkler,Z+G,P+V} for further details of
this perturbative expansion.

We also make the assumption that in this basis, the saddle point
value of $Q$ is independent of the Landau indices and $x_0$: this
approximation is analogous to the C-number approximation (CNA)
introduced by Zhang and Gerhardts~\cite{Z+G} for the self-energy
matrix in the presence of modulation, and is valid as long as the
magnetic field is not too high. In order to go beyond the CNA, one
would need to generalize the saddle point $Q$ to include a matrix
structure in the space of Landau indices and a dependence on
$x_0$: such a task is of interest for future work but beyond the
scope of this paper.

The saddle point equation (\ref{sad1}) may now be written
\begin{eqnarray}
\frac{i \hbar Q}{2\tau}  = \Gamma_0^2\sum_n \frac{1}{a}
\int_0^a dx_0 \frac{1}{E-E_n(x_0)-i\delta \Lambda -
\frac{i \hbar Q}{2\tau}}\, ,
\label{sad3}
\end{eqnarray}
where $\Gamma_0$ is the width of the
Landau level in the absence of modulation and

\begin{eqnarray}
\Gamma_0^2=\frac{1}{2\pi} \hbar\omega_c \frac{\hbar}{\tau}.
\end{eqnarray}

The saddle point equation (\ref{sad3}) coincides
 with the self-energy equation derived in the self-consistent Born
 approximation (SCBA) by Zhang and Gerhardts\cite{Z+G}, under the replacement $i\hbar
Q^{1,2}/(2\tau) \to \Sigma^{R,A}$, where indices refer to
advanced/retarded space and $\Sigma$ denotes the self-energy. The
solution for $Q$ is then of the form $i Q = e_0+ i\Lambda \rho_0$,
where $\rho_0$ is proportional to the density of states. In the
absence of modulation, the density of states reduces to that
determined by Ando and co-workers\cite{Ando} within the SCBA (as
shown by Pruisken\cite{Pruisken}).

Having identified the saddle point value $Q$, we proceed by
performing a gradient expansion of the Lagrangian (\ref{Trln})
around the saddle point.  We follow very closely the calculation
of Pruisken\cite{Pruisken}, although we use the supersymmetric
rather than the replica formulation. We use the representation

\begin{eqnarray*}
Q({\bf r}) = T({\bf r}) P({\bf r}) T^{-1}({\bf r}),
\end{eqnarray*}
where the $P$ fields are diagonal in retarded/advanced space and
represent the massive modes. The procedure is to integrate over
$P$ and $T$ separately. To do so, we split the fields of
integration,

\begin{eqnarray*}
\int D Q = \int DP \int DT \exp({\rm Str}\, {\rm ln}[I[P]]),
\end{eqnarray*}
in the process acquiring the associated Jacobian
$I[P]$ (as discussed in Ref~.\cite{Pruisken}). ${\cal L}$ becomes

\begin{eqnarray*}
{\cal L}[P,T] &=& \int dr
\left\{
-{\rm Str}\,{\rm ln}[I[P]] +\frac{\hbar \pi\nu}{8\tau}
{\rm Str}P^2
\right. \\
&&\left.
-\frac{1}{2}{\rm Str}\,{\rm
  ln}\left[
i\left(
E-T^{-1} \pi_{\mu}\pi^{\mu} T -U_0\cos(qx)-\frac{i \hbar
  P}{2\tau}\right)
\right]\right\}.
\end{eqnarray*}
To integrate over the $P$ modes, wsplit ${\cal L}$ into two parts:
\begin{eqnarray*}
{\cal L}_0[P] &=& \int dr\left\{
-{\rm Str}\,{\rm ln}[I[P]] +\frac{\hbar \pi\nu}{8\tau}
{\rm Str}P^2
-\frac{1}{2}{\rm Str}\,{\rm
  ln}\left[E-\pi_{\mu}\pi^{\mu} -U_0\cos(qx)-\frac{i \hbar P}{2\tau}\right]
\right\}, \\
\delta {\cal L}[P,T] &=& {\cal L}[P,T]-{\cal L}_0[P].
\end{eqnarray*}

Integration over the $P$ fields then proceeds by cumulant-expanding
$\delta {\cal L}[P,T]$ with respect to ${\cal L}_0[P]$.
In turn, $\delta {\cal L}[P,T]$ is
computed by a gradient expansion of the
Str ln term up to second order in the combination
$D_{\mu}\equiv T^{-1} \nabla_{\mu} T$. The propagators for
the latter expansion are of the form

\begin{eqnarray*}
g({\bf r},{\bf r}') = <{\bf r}|\left(E- \pi_{\mu}\pi^{\pi}
-U_0\cos(qx)-\frac{i \hbar P}{2\tau}({\bf r})\right)^{-1}|{\bf r}'>,
\end{eqnarray*}
weighted with respect to ${\cal L}_0[P]$. A typical second-order term is

\begin{eqnarray*}
{\cal L}_{\rm typ} = \int dr \int dr'
{\rm Str}[D_{\mu}({\bf r})]
[D_{\nu}({\bf r}')]
\langle g({\bf r}',{\bf r})\pi^\mu g({\bf r},{\bf
  r}')
\pi^\nu\rangle,
\end{eqnarray*}
where the averaging is over $P$ with respect to ${\cal L}_0$. At
this point, we now exploit the assumption that the $T({\bf r})$
matrices vary in space more slowly than the modulation. Thus the
averages of the products of the Green functions are short ranged
with respect to the $T$ matrices, and are translationally
invariant after averaging over one cycle of the modulation. This
allows us to
 perform a gradient expansion in the propagator averages:

\begin{eqnarray*}
\langle g({\bf r}',{\bf r})\pi^\mu g({\bf r},{\bf
  r}')
\pi^\nu\rangle &=& K({\bf r}-{\bf r}')
\\
&=& K^{(0)}\delta({\bf r}-{\bf r}') + K^{(2)}\delta({\bf r}-{\bf r}')
\nabla_{\mu} \nabla_{\nu}+\ldots,
\end{eqnarray*}
here the averaging includes that over a cycle of the modulation. A
series of Ward identities may now be used as in
Ref.~\cite{Pruisken} to simplify the resulting expressions up to
second order in $D_{\mu}$. The final Lagrangian is then of the
form displayed in Eq.~(\ref{diffaction}), where $Q=T \Lambda
T^{-1}$.

Again, we have the standard form of a nonlinear $\sigma$ model
with a topological term, although with the anisotropy
$\sigma^0_{xx}\neq\sigma^0_{yy}$. Again, $Q$ satisfies $Q^2=1$ and
the symmetries $Q=\bar{Q}= K Q^{\dagger} K$.  Also the bare
coefficients, $\sigma^0_{ij}$, are the components of the
conductivity tensor at zero temperature in units of $e^2/h$,
according to the formulas
\begin{eqnarray}
\sigma^0_{xx,yy} &=& -\frac{\hbar^2}{m}
\int dr'
\langle \pi_{x,y}(g^1({\bf r},{\bf r}')-g^2({\bf r},{\bf
  r}'))
\pi_{x,y}(g^1({\bf r}',{\bf r})-g^2({\bf r}',{\bf
  r}))\rangle,
\nonumber \\
\sigma^0_{xy} &=& \sigma^{0,I}_{xy}+ \sigma^{0,II}_{xy}, \nonumber \\
\sigma^{0,II}_{xy} &=& \frac{i\hbar c}{e}
\left\{
\int_{-\infty}^{E}dE^1 \frac{\partial }{\partial B^1}\langle
g^1({\bf r},{\bf r})\rangle
- \int_{-\infty}^{E}dE^2 \frac{\partial }{\partial B^2}\langle
g^2({\bf r},{\bf r}))
\rangle\right\}, \nonumber\\
\sigma^{0,I}_{xy} &=& \frac{\hbar^2}{m}
\int dr'\left\{
\langle \pi_x g^1({\bf r},{\bf r}') \pi_y g^2({\bf r}',{\bf r})\rangle  -
\langle \pi_y g^1({\bf r},{\bf r}') \pi_x g^2({\bf r}',{\bf r})\rangle
\right\},
\label{condtensor2}
\end{eqnarray}
here the averaging is over $P$ as well as over one cycle of the
modulation with respect to ${\bf r}$. The propagator averages may
now be evaluated by their saddle point values with respect to $P$.

The bare coefficients $\sigma^0_{ij}$ are the components of the
conductivity tensor, in units of $e^2/h$, calculated in the SCBA
at zero temperature. In the absence of modulation, the expressions
(\ref{condtensor2}) for the bare conductivity then reduce to those
computed in the SCBA by Ando and co-workers~\cite{Ando}. In the
presence of modulation, they reduce to the conductivity tensor
computed in the SCBA by Zhang and Gerhardts~\cite{Z+G}:
$\sigma_{ij}^0 = \sigma_{ij}^{\rm qu}$ (the equivalence of may be
seen by following the working of Streda\cite{Streda2}). Therefore,
in contrast to those derived via the ballistic $\sigma$ model, the
bare conductivities in the Lagrangian (\ref{diffaction}) represent
the values calculated by a fully quantum-mechanical approach.

%**************************************

%* Scaling                            *

%**************************************

\section{Weak localization and Quantum Hall effect}

\label{sec:scaling}

Having derived the final form of the Lagrangian
(\ref{diffaction}), we are now in a position to use it to
calculate weak localization corrections to the conductivity
tensor. The effect of the corrections may be expressed as a
scaling of the conductivity tensor under changes of length scale.
A perturbative treatment of the Lagrangian (\ref{diffaction}) is
valid in the limit of large longitudinal conductivity
($\sigma^0_{xx} \gg 1$). Within perturbation theory, however, the
Hall conductivity is not renormalized by weak localization
corrections. Instead, the effect of the topological term in the
Lagrangian may only be made apparent through a nonperturbative
analysis. Such an analysis
 has been
conjectured\cite{Levine,Khmelnitskii,Pruis2} in the form of a
two-parameter renormalization-group procedure.  The two parameters
are the longitudinal and Hall conductivities which follow coupled
scaling equations with respect to changes of length scale. The
derivation of the scaling equations presented in
Refs.~\cite{Levine,Pruis2} is based on approximating
configurations of the field-theory parameter as a gas of
instantons. While the validity of this derivation is being
vigorously debated (see, e.g., Refs.~\cite{Zirnbauer,Weid+Z}), it
has remained a valuable guide to experimental\cite{Wei} and
numerical\cite{Huckestein} data for a number of years. In the
following, we take a pragmatic approach by not contesting the
validity of the two-parameter scaling and its relation to the
proposed field theory. Instead, we assume its validity and explore
how it may be generalized to take account of the periodic
potential.

\subsection{Weak localization}
First we subject the Lagrangian to a perturbative RG analysis. The
perturbation theory is valid in the limit of $\sigma^0_{xx} \gg
1$. The topological term in the Lagrangian does not contribute at
any perturbative order and hence $\sigma_{xy}$ is unrenormalized.
For an unmodulated system, $\sigma_{xx}=\sigma_{yy}$ and the
perturbative RG procedure has been explained in detail, e.g., in
Ref.~\cite{Efe97}.

In the presence of modulation, the $\sigma$ model is slightly
nonstandard due to the anisotropy of the longitudinal
conductivities, $\sigma_{xx}\neq \sigma_{yy}$. However, quantum
interference processes in a diffusive, anisotropic conductor have
been considered before, for example by W\"olfle and
Bhatt\cite{Woe84}, where the origin of the anisotropy was
envisaged as due to a difference in effective electron masses in
the two directions. In the latter paper, a diffusive $\sigma$
model was derived with the same form as Eq.~(\ref{diffaction}),
although without the topological term.

When $\sigma_{xx}\neq \sigma_{yy}$, the two parameters, $\sigma_{xx}$
and $\sigma_{yy}$, follow
coupled flow equations under changes of length scale. While it is
straightforward to derive and solve the two flow equations, it is also
instructive to follow a different strategy whereby we perform a scale
transformation after which the
Lagrangian (\ref{diffaction})  maps to an isotropic form: we scale

\begin{eqnarray}
x' = x(\sigma^0_{yy}/\sigma^0_{xx})^{1/4}, \qquad
y' = y(\sigma^0_{xx}/\sigma^0_{yy})^{1/4},
\label{scale}
\end{eqnarray}
under which the Lagrangian (\ref{diffaction}) transforms to an
isotropic $\sigma$ model:

\begin{eqnarray}
{\cal L}[Q] = \frac{1}{16}
\int dr' {\rm Str}\left\{ \tilde{\sigma}^0
(\nabla Q)^2
-\sigma^0_{xy} \tau_3
Q[\nabla_{x'} Q,\nabla_{y'} Q]\right\},
\label{Pruisken2}
\end{eqnarray}
where $\tilde{\sigma}^0 \equiv \sqrt{\sigma_{xx}^0 \sigma_{yy}^0}$. The
perturbative scaling for $\tilde{\sigma}$ may now be derived in the
standard way\cite{Efe97}: the flow equation is (to leading order)

\begin{eqnarray*}
\frac{d \tilde{\sigma}}{d\log{L}}&=&-\frac{1}{2\pi^2\tilde{\sigma}}.
\end{eqnarray*}

By integrating the flow equation from microscopic to macroscopic
length scales (of the order of the system size $L$), we find the
conductivity

\begin{eqnarray}
\tilde{\sigma}(L)=\tilde{\sigma}^0
\left(1-\frac{1}{\pi^2}
\frac{1}
{(\tilde{\sigma}^0)^2}
\log\frac{L}{\ell} \right)^{1/2}.
\label{pert}
\end{eqnarray}

Using the fact that the ratios $\sigma_{xx,yy}/\tilde{\sigma}$ are
invariant under the scaling, we may recover $\sigma_{xx,yy}(L)$
from $\tilde{\sigma}(L)$ as follows:

\begin{eqnarray}
\sigma_{xx}(L) &=& \tilde{\sigma}(L) \sqrt{\sigma_{xx}^0
/\sigma_{yy}^0}, \nonumber\\
\sigma_{yy}(L) &=& \tilde{\sigma}(L) \sqrt{\sigma_{yy}^0
/\sigma_{xx}^0}.
\label{sigtil}
\end{eqnarray}

We remark that the perturbative scaling derived from a field theory as above
in the presence of anistropy
in the conductivities in the $x$ and $y$ directions (due to an
anisotropy in the electron mass) has been confirmed by direct
diagrammatics (in the orthogonal ensemble)
by W\"olfle and Bhatt\cite{Woe84}.

Equation (\ref{pert}) shows how weak localization corrections
affect the longitudinal conductivity within perturbation theory.
We can see how the conventional scaling applies to
$\tilde{\sigma}$ in the presence of the periodic potential. By
inputting the SCBA values for $\sigma^0_{xx}$ and $\sigma^0_{yy}$
weak localization corrections are obtained, according to
Eq.~(\ref{pert}) they depend on {\it both} bare values
$\sigma^0_{xx}$ and $\sigma^0_{yy}$. Corrections to $\sigma_{xx}$,
for example, are influenced by the strong oscillatory behavior of
$\sigma_{yy}^0$. Upon inverting the conductivity tensor to find
the resistivity tensor, the weak localization corrections will
therefore give rise to additional oscillatory corrections to the
resistivity tensor as a function of the magnetic field. As
mentioned earlier, strong Weiss oscillations in $\rho_{xx}$ are
accompanied by weak out-of-phase oscillations in $\rho_{yy}$ as
long as quantum interference effects may be neglected. It is
interesting to notice that weak localization corrections to
$\rho_{yy}$ oscillate in phase with the (dominant) oscillations in
$\rho_{xx}$.

In practice\cite{Wei}, rather than changing the system size $L$,
one varies the temperature to change the effective system size,
which is given by the diffusion length, $(\hbar D/(k_B T))^{1/2}$.
The perturbative result (\ref{pert}) is only valid when the
corrections are much smaller than the bare conductivities, and
hence may be difficult to verify experimentally. A potentially
more promising strategy is to illustrate the effect of weak
localization corrections when their contribution is relatively
large such as is the case for the IQHE.

\subsection{Quantum Hall Effect}

We now generalize from the perturbative analysis to discuss
the nonperturbative scaling that would follow from the Lagrangian (\ref{diffaction}). Again, we are able to make the scale transformation (\ref{scale}) as above, which maps the Lagrangian to the isotropic version, Eq.~(\ref{Pruisken2}).
The coupled flow of the two coupling constants, $\tilde\sigma$ and $\sigma_{xy}$, may written in a general form,

\begin{eqnarray}
\frac{d\tilde\sigma}{d\log{L}} = \tilde{\beta}(\tilde\sigma,\sigma_{xy})
\qquad
\frac{d\sigma_{xy}}{d\log{L}} = \beta_{xy}(\tilde\sigma,\sigma_{xy}).
\label{genscale}
\end{eqnarray}
The above beta functions, $\tilde\beta$ and $\beta_{xy}$, would
take precisely the same form as the beta-functions that describe
the flow of $\sigma_{xx}$ and $\sigma_{xy}$ in the equivalent,
unmodulated system.

The starting point of the flow is determined by the bare
conductivity tensor $\tilde{\sigma}_{ij}$. In general, the coupled
flow equations (\ref{genscale}) for $\tilde{\sigma}$ and
$\sigma_{xy}$ need to be integrated up to length scales of the
system size. The values of $\sigma_{xx}(L)$ and $\sigma_{yy}(L)$
may then be recovered from $\tilde{\sigma}(L)$
 by Eqs.~(\ref{sigtil}).
As a consequence of Eq.~(\ref{sigtil}), we see
that
the ratio $\sigma_{xx}(L)/\sigma_{yy}(L)$ remains constant under the scaling.
This conclusion is furthermore independent of the form of the
beta-functions, $\tilde\beta$ and $\beta_{xy}$, in Eq.~({\ref{genscale}}).

\begin{figure}[tbp]
\begin{center}
\epsfig{file=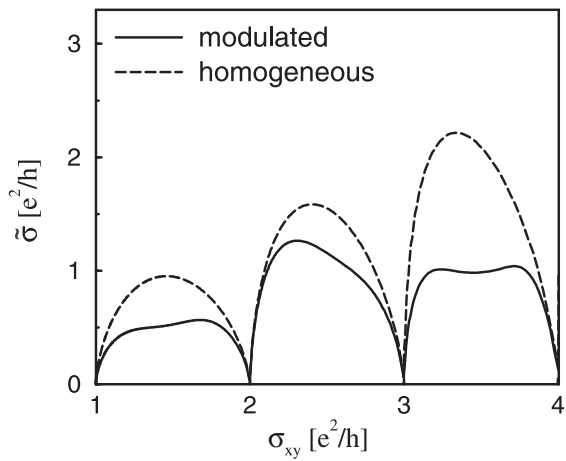,height=6cm} \caption{\label{fig:SCALING}
Plot of bare conductivities, providing a starting point for
scaling, for a modulated vs homogeneous system. Parameters:
$a=$140 nm, $U_0=$0.2 meV, $\Gamma_0=$0.048$\sqrt{B[T]}$ meV and
$n_{el}=$1$\times$10$^{11}$ cm$^2$.}
\end{center}
\end{figure}

In the absence of modulation, the SCBA
values\cite{Ando} of the bare conductivities
correspond to an approximate semicircular dependence
of $\sigma^0_{xx}$ on $\sigma^0_{xy}$. In the presence
of modulation, the dependance of $\tilde{\sigma}^0$ on $\sigma^0_{xy}$
is modified  from the semicircle in a complicated way; a typical
dependence, calculated using the scheme proposed in Ref.~\cite{Z+G},
is shown in Fig.~\ref{fig:SCALING}.

%***************************************************

%* Figure: SCALING                                   *

%***************************************************

\begin{figure}[tbp]
\begin{center}
\epsfig{file=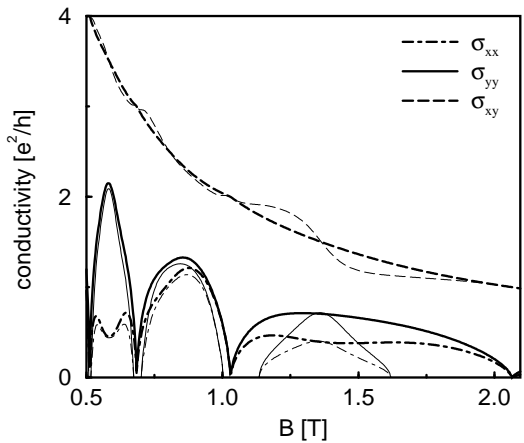,height=6cm}
\caption{\label{fig:condscaling} Evolution of the conductivity tensor
  for the modulated system with system size $L$ under the two-parameter
  scaling. The thick curves are the bare conductivities ($L=L_0$), and the thin curves are
  the scaled conductivities for  $\log(L/L_0)=2$.
Parameters as in Fig.~\ref{fig:SCALING}.}
\end{center}
\end{figure}

The corresponding dependency of the bare conductivity tensor on
magnetic field for the same parameters is part of
Fig.~\ref{fig:condscaling} below (thick curves). We remark that
for the peaks centered around $B$=1.5 T one may already expect
vertex corrections to become effective, although they are not
included
 in the CNA approach that we have employed.
 Such corrections have the general tendency\cite{Manolescu,Gross}
to enhance $\sigma^0_{yy}$
in comparison to $\sigma^0_{xx}$.

In order to provide an illustration of the IQHE in the modulated
sample, it is necessary to assume a particular form of the scaling
equations~({\ref{genscale}}). The following scaling
equations\cite{Levine,Pruis2,P+Girvin,Huckestein} were derived
(originally in the replica formulation) within a dilute gas
approximation of the instantonic configurations of the $Q$ matrix
from the Lagrangian (\ref{diffaction}):
\begin{eqnarray}
\frac{d \tilde{\sigma}}{d\log{L}}&=&-\frac{1}{2\pi^2\tilde{\sigma}}
-\tilde{\sigma}^3D_c\cos(2\pi\sigma_{xy})\exp(-2\pi\tilde{\sigma}),
\nonumber\\
\frac{d\sigma_{xy}}{d \log{L}}&=&
-\tilde{\sigma}^3 D_c \sin(2\pi\sigma_{xy})\exp(-2\pi\tilde{\sigma}).
\label{twopar}
\end{eqnarray}
The dimensionless constant $D_c$ is of order unity and is related
to the density of instantons. It may be seen immediately from the
form of Eq.~(\ref{twopar}), that along the lines
$\sigma_{xy}=(n+1/2)$, where $n$ is an integer, the Hall
conductance is unrenormalized. It may also been seen that the
points $(\sigma_{xy},\tilde{\sigma})=(n,0)$, for integer $n$, are
(attractive) fixed points. Upon scaling the system from
microscopic to macroscopic length scales, the coordinates
$(\sigma_{xy},\tilde{\sigma})$ scale from the bare nonuniversal
values towards the quantized values $(n,0)$, for integer $n$. This
tendency reflects the quantization of the Hall conductivity under
scaling at low temperatures.

\begin{figure}[tbp]
\begin{center}
\epsfig{file=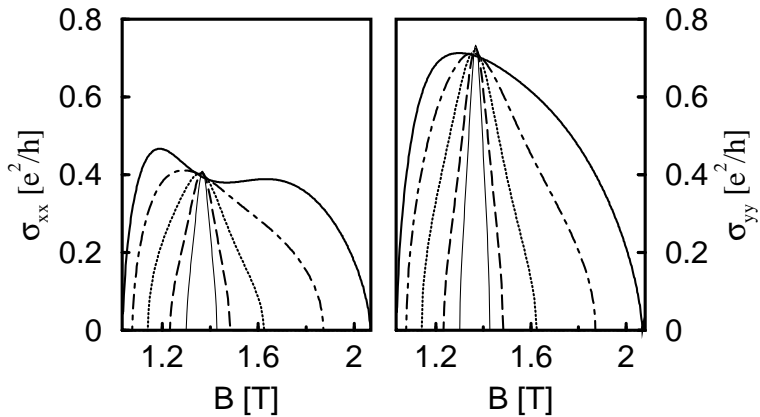,height=6cm}
\caption{\label{fig:FOCUSCONDSCALING} Focus on the evolution of
the longitudinal conductivities with system size L under the
two-parameter scaling. Parameters as in
Fig.~\ref{fig:condscaling}, and $\log(L/L_0)=0$--4, from the
broader to the narrower peaks.}
\end{center}
\end{figure}

As mentioned before, the validity of the
approximations underlying Eq.~({\ref{twopar}})
has long been the subject of debate\cite{Weid+Z,Zirnbauer}.
Keeping this in mind we use these equations to provide an illustration
of the operation of the IQHE in the modulated
system in Figs.~\ref{fig:condscaling} and \ref{fig:FOCUSCONDSCALING}.

%***************************************************

%* Figure: CONDSCALING                        *

%***************************************************

Fig.~\ref{fig:condscaling} shows the evolution of the conductivity
for the modulated system under the two-parameter scaling. We see
how the Hall conductivity becomes quantized under scaling in the
modulated system. Between the plateaus in the Hall conductivity,
the longitudinal conductivities develop peaks under scaling of
differing heights, due to the anisotropy in the system.
Fig.~\ref{fig:FOCUSCONDSCALING} shows in more detail how these
peaks in the longitudinal conductivities develop, while the ratio
between the two conductivities remains constant under scaling.

%***************************************************
%* Figure: FOCUSCONDSCALING                        *
%***************************************************

We remark that the behavior described above should be within
current experimental capabilities: for example, Geim {\em et
al}.\cite{Geim} have studied the high magnetic field ($R_c \ll a$)
regime, although they kept the temperature high to avoid the
effect of quantum interference processes. Bagwell {\em et
al}.\cite{Bagwell} have also studied the IQHE in modulated
samples; in their work however they have not focused on the regime
of a weak periodic potential or effects that are independent of
the device boundaries.

%*************************************************

%* Summary                                       *

%*************************************************

\section{Summary and Discussion}

\label{sec:summary}

In this paper we have considered transport properties of a
disordered conductor in a periodic potential and strong magnetic
field. We focused on the contribution of quantum interference
processes, whose influence at high fields is missing from previous
approaches despite being responsible for a whole class of
phenomena. For example, they lead to weak localization corrections
to the conductivity and the operation of the IQHE at high magnetic
fields. To this end, we introduced a field-theory approach, which
is well established in the study of unmodulated disordered
conductors.

The effective Lagrangian of the field theory takes the form of a
nonlinear $\sigma$ model, which describes the interaction of
diffusion modes on large length scales. The presence of the strong
magnetic field leads to an extra, topological term in the
Lagrangian. The form of the Lagrangian is the same as for
unmodulated systems, except for an anisotropy in the coefficients
corresponding to the bare longitudinal conductivities in the $x$
and $y$ directions.

We provided two different routes to deriving the Lagrangian. The
first route was via a so-called "ballistic" $\sigma$ model, and
demonstrated how the results of the quasiclassical
approach\cite{Beenakker} may be recovered within the field-theory
formalism. The drawback of this route, in common with the
quasiclassical approximation, is that it neglects the
renormalization of single-particle properies by Landau
quantization. Consequently the resulting Lagrangian was too
approximate to be useful in determining weak localization
properties at low temperatures. The second route improved on this
situation by including the effects of Landau quantization. It
bypasses a derivation of a ballistic $\sigma$ model and instead
follows more closely the original derivation of
Pruisken\cite{Pruisken} for unmodulated systems. Indeed, while the
first route contained more parallels with the quasiclassical
approach, the second route contained more parallels with the
quantum-mechanical approach\cite{Z+G} for the bare conductivity
tensor.

Having derived the effective Lagrangian, we showed how it leads to
the scaling of the conductivity tensor under changes of length
scale (and hence temperature). A perturbative
renormalization-group analysis of the free energy leads to a
generalization, to modulated systems,
 of one-parameter scaling for
the longitudinal conductivities. Perturbatively, the Hall
conductivity is unrenormalized. In the regime of Weiss
oscillations\cite{Weiss}, weak localization corrections give rise
to an additional oscillatory dependence of the longitudinal
conductivities as a function of the magnetic field. Due to their
smallness, these corrections may, however, be hard to detect
experimentally.

In order to describe the IQHE, whereby the Hall conductivity
becomes quantized at low temperatures, a nonperturbative analysis
of the Lagrangian is necessary. This is provided by the conjecture
of the two-parameter scaling for unmodulated systems. Assuming the
validity of this conjecture and the underlying instanton gas
approximation, we have shown how the two-parameter scaling  may be
generalized to the case of the modulated system. This has allowed
us to illustrate the evolution of the resistivity tensor under the
IQHE for parameters that are realistic for experiments. We find,
for example, that the ratio of the two longitudinal components of
the conductivity remains constant under scaling (an observation
that does not itself depend on the assumptions used to derive the
flow equations).
 While the Hall conductivity still becomes quantized under
scaling in the
modulated system, between the plateaus the longitudinal conductivities
develop peaks of differing heights due to the anisotropy in the system.

There are several directions in which our analysis may be
generalized. A simple generalization is to consider a
periodic magnetic field, rather than potential, a situation which
leads to a similar phenomenology\cite{Xue,Menne,Gerhardts}. The
derivation of the field theory for this case follows very similar
lines to the case of a periodic potential, with similar results.

Another direction in which the analysis may be extended
straightforwardly is to the study of the low magnetic field
regime, in which a positive magnetoresistance has been
observed\cite{Weiss,Beton} and explained within a quasiclassical
approach\cite{Beton,Streda,Menne}. As long as $\omega_c\tau \ll
1$, one may derive the field theory for this regime according to
the first route of Sec.~\ref{sec:sigma}, neglecting quantization;
the coefficients of the field theory then coincide with the
components of the conductivity tensor derived within the Boltzmann
approach. The weak localization to the conductivity (or related
quantities such as the dephasing time\cite{Wang}) may then be
calculated using the Lagrangian for either the orthogonal or the
unitary ensemble, or in the crossover between the pure symmetry
classes.

A less straightforward generalization is to improve on the CNA
approximation of the quantum-mechanical approach and its analog in
the field-theory formalism. According to this approximation, the
self-energy matrix (and the saddle point solution of the $Q$
matrix) are assumed to have a trivial structure in the space of
Landau indices and coordinate $x_0$. While this simplifies the
analysis considerably, we have seen already how this approximation
breaks down for very high magnetic fields, such that the cyclotron
radius is much less than the period of modulation, $R_c \ll a$.
Improvement on the CNA is necessary, not only for such high
fields, but also to study models of disorder with long-range
correlations, for which vertex corrections are not negligible.
Such models have been analyzed in the quasiclassical
approach\cite{M+Wolf} and shown to represent experimental data
more closely in certain respects.

Very recently it has been shown how to improve on the
CNA\cite{Manolescu,Gross} within the quantum-mechanical approach,
so as to analyze models at high magnetic fields or with long-range
disorder correlations. In the field theory formalism, improvement
on the CNA approximation would require enlarging the space of the
$Q$ matrix even further to include the additional matrix structure
in the space of Landau indices, as well as a dependence on $x_0$.
Inclusion of such a structure within the $Q$ field is to our
knowledge a novel direction to pursue, although this task is left
as a future project.

A further area to explore is the case of a strong periodic
potential, for which the first-order perturbative expansion in the
potential that we use is no longer valid. While an analytical
approach would be very difficult, a numerical treatment would be
better suited to this regime. For a strong enough potential that
$U_0$ is of the order of $\omega_c$, the smearing of the Landau
levels is so great that they can no longer be individually
resolved even at low temperatures. This leads to a quenching of
the Shubnikov--de Haas oscillations (as noted by Beton {\em et
al}.\cite{Beton}) and hence one would expect the quantum Hall
effect to be destroyed.

\acknowledgements

We would like to thank B. Huckestein and M. Langenbuch for helpful discussions during the course of this work.
The authors gratefully acknowledge the financial support of Trinity College, Cambridge (D. T.-S.), the {\it Graduiertenkolleg} 384 (G.S.), the {\it Schwerpunktprogramm ``Quanten Hall Systeme''} and the {\it Sonderforschungsbereich} 491.


\begin{references}
\bibitem{Weiss} D. Weiss, K. v. Klitzing, K. Ploog and G. Weimann,
  Europhys. Lett. {\bf 8}, 179 (1989); also in {\em High Magnetic
  Fields in Semiconductor Physics II}, vol. 87 of {\em Springer Series
  in Solid-State Sciences}, ed. G. Landwehr, Springer-Verlag: Berlin (1989).
\bibitem{Beton} P. H. Beton, E. S. Alves, P. C. Main, L. Eaves,
  M. W. Dellow, M. Henini, O. H. Hughes, S. P. Beaumont and
  C. D. W. Wilkinson, Phys. Rev. B {\bf 42}, 9229 (1990).
\bibitem{Streda} P. Streda and A. H. MacDonald, Phys. Rev. B {\bf 41},
  11892 (1990); P. Streda, J. Kucera and J. van de Konijnenberg,
  Physica Scripta {\bf T39}, 162 (1991); J. Kucera, P. Streda and
  R. R. Gerhardts, Phys. Rev. B {\bf 55}, 14439 (1997).
\bibitem{Menne} R. Menne and R. R. Gerhardts, Phys. Rev. B {\bf 57},
  1707 (1998).
\bibitem{GWK} R.R. Gerhardts, D. Weiss and K. v. Klitzing,
  Phys. Rev. Lett. {\bf 62}, 1173 (1989).
\bibitem{Winkler} R. W. Winkler, J. P. Kotthaus and K. Ploog,
  Phys. Rev. Lett. {\bf 62}, 1177 (1989).
\bibitem{Z+G} C. Zhang and R.R. Gerhardts, Phys. Rev. B
{\bf 41}, 12850 (1990).
\bibitem{P+V} P. Vasilopoulos and F. M. Peeters, Phys. Rev. Lett. {\bf
  63}, 2120 (1989); F. M. Peeters and P. Vasilopoulos, Phys. Rev. B {\bf 46}, 4667 (1992).
\bibitem{Klitzing} K. v. Klitzing, G. Dorda and M. Pepper,
  Phys. Rev. Lett. {\bf 45}, 494 (1980).
\bibitem{Manolescu} A. Manolescu, R.R. Gerhardts, M. Suhrke and U. R\"ossler,
Phys. Rev. B {\bf 63}, 115322 (2001).
\bibitem{Gross} J. Gross and R. R. Gerhardts, Physica B {\bf 298}, 83 (2001).
\bibitem{Beenakker} C. W. J. Beenakker, Phys. Rev. Lett. 62, 2020
  (1989).
\bibitem{M+Wolf} A. D. Mirlin and P. W\"olfle, Phys. Rev. B {\bf 58},
  12986 (1998).
\bibitem{Geim} A. K. Geim, R. Taboryski, A. Kristensen, S. V. Dubonos
  and P. E. Lindelof, Phys. Rev. B {\bf 46}, 4324 (1992).
\bibitem{Bagwell} P. F. Bagwell, S. L. Park, A. Yen, D. A. Antoniadis, H. I. Smith, T. P. Orlando and M. A. Kastner, Phys. Rev. B {\bf 45}, 9214 (1992).
\bibitem{Milton} B. Milton, C. J. Emeleus, K. Lister, J. H. Davies and
  A. R. Long, Physica E {\bf 6}, 555 (2000).
\bibitem{Wegner} F. Wegner, Z. Phys. B {\bf 35}, 207 (1979).
\bibitem{ELK} K. B. Efetov, A. I. Larkin and D. E. Khmelnitskii,
Zh. Eksp. Teor. Fiz. {\bf 79}, 1120 (1980) [JETP {\bf 52}, 568
(1980)].
\bibitem{Efe97} K. B. Efetov, Adv. Phys. {\bf 32}, 53 (1983); K. B.
Efetov, {\it Supersymmetry in Disorder and Chaos}, Cambridge University
Press, New York (1997).
\bibitem{Pruisken}  A. M. M. Pruisken, Nucl. Phys. B {\bf 235}, 277
  (1984)
\bibitem{Levine} H. Levine, S. B. Libby and A. M. M. Pruisken,
  Nucl. Phys. B240 [FS12], 30; 49; 71 (1984).
\bibitem{Ando} T. Ando and Y. Uemura, J. Phys. Soc. Jpn. {\bf 36},
  359 (1974); T. Ando, J. Phys. Soc. Jpn. {\bf 36}, 1521 (1974); {\bf
  37}, 622 (1974); {\bf 37}, 1233 (1974);
T. Ando, A. Fowler and F. Stern, Rev. Mod. Phys. {\bf
  45}, 437 (1982).
\bibitem{Altshuler} B. L. Altshuler, A. G. Aronov, D. E. Khmelnitskii
  and A. I. Larkin, in {\em Quantum Theory of Solids}, ed. I. M. Lifshitz
  (MIR Publishers:Moscow) (1983).
\bibitem{Bergmann} G. Bergmann, Phys. Rep. {\bf 101}, 1 (1984).
\bibitem{Chakravarty} S. Chakravarty and A. Schmid, Phys. Rep. {\bf
    140}, 193 (1986).
\bibitem{Belitz} D. Belitz and T. R. Kirkpatrick, Rev. Mod. Phys. {\bf
    66}, 261 (1994).
\bibitem{GLK} L. P. Gorkov, A. I. Larkin and D. E. Khmelnitskii,
  Pis'ma Zh. Eksp. Teor. Fiz. {\bf 30}, 248 (1979) [JETP Lett. {\bf
    30}, 228 (1979)].
\bibitem{Hikami} S. Hikami, A. I. Larkin and Y. Nagaoka,
  Prog. Theor. Phys. {\bf 63}, 707 (1980);
S. Hikami, Phys. Rev. B {\bf 24}, 2671 (1981).
\bibitem{Shelankov} J. Rammer and A. L. Shelankov, Phys. Rev. B {\bf
    36}, 3135 (1987).
\bibitem{Wang} X.-B. Wang, Phys. Rev. B {\bf 65}, 115303 (2002).
\bibitem{Khmelnitskii} D. E. Khmelnitskii, Pis'ma
  Zh. Eksp. Teor. Fiz. {\bf 38}, 454 (1983) [JETP Lett. {\bf 38}, 552 (1983)].
\bibitem{Pruis2} A. M. M. Pruisken, Phys. Rev. B {\bf 32}, 2636 (1985).
\bibitem{P+Girvin} A. M. M. Pruisken, in {\em The Quantum Hall Effect},
ed. R. E. Prange and S. M. Girvin, Springer:Berlin (1987).
\bibitem{Weid+Z} H. A. Weidenmuller and M. R. Zirnbauer, Nucl. Phys. B
  {\bf 305 [FS25]}, 339 (1988).
\bibitem{Zirnbauer} M. R. Zirnbauer, Ann. Phys. (Leipzig) {\bf 3}, 513 (1994).
\bibitem{Huckestein} B. Huckestein, Rev. Mod. Phys. {\bf 67}, 357 (1995).
\bibitem{Wei} H. P. Wei, D. C. Tsui and A. M. M. Pruisken,
  Phys. Rev. B {\bf 33}, 1488 (1985).
\bibitem{Abr72} M. Abramowitz and I. A. Stegun,
{\em Handbook of Mathematical Functions}, Dover Publications, Inc.:
New York (1972).
\bibitem{Muz95} B. A. Muzykantskii and D. E. Khmelnitskii, Pis'ma Zh. Eksp. Teor. Fiz. {\bf 62}, 68 (1995) [JETP Lett. {\bf 62}, 76 (1995)].
\bibitem{Andreev} A. V. Andreev, B. D. Simons, O. Agam and B. L. Altshuler,
Nucl. Phys. B {\bf 482}, 536 (1996).
\bibitem{Woe84} P. W\"olfle and R.N. Bhatt, Phys. Rev. B {\bf 30},
  3542 (1984).
\bibitem{Aronov} A. G. Aronov, A. D. Mirlin and P. W\"{o}lfle, Phys. Rev. B
{\bf 49}, 16 609 (1994).
\bibitem{TE} D. Taras-Semchuk and K. B. Efetov, Phys. Rev. B {\bf 64},
  115301 (2001).
\bibitem{note:semi} The semiclassical approximation referred to here
  differs from the quasiclassical approximation associated with the
  Boltzmann equation approach of Ref.~\cite{Beenakker}.
\bibitem{Xue} D. P. Xue and G. Xiao, Phys. Rev. B {\bf 45}, 5986 (1992).
\bibitem{Gerhardts} R. R. Gerhardts, Phys. Rev. B {\bf 53}, 11064 (1996).
\bibitem{Streda2} P. Streda, J. Phys. C {\bf 15}, L717 (1982).
\end{references}
\end{document}